\documentclass[a4paper]{jpconf}
\usepackage{graphicx}
\usepackage[usenames,dvipsnames]{color}
\usepackage[english]{babel} 
\usepackage{amsmath}
\usepackage{subfig}
\usepackage{amsfonts}
\usepackage{cite}
\usepackage{amssymb}
\usepackage{float}
\usepackage[T1]{fontenc}
\usepackage{textcomp}
\usepackage{geometry}
\usepackage{epstopdf}
\usepackage[labelfont=bf]{caption}

\begin{document}
\title{Emergence of correlations in highly biased Consensus Models in seed initial configuration}

\author{M Di Vece$^1$ and F Corberi$^{1,2}$}

\address{$^1$ Dipartimento di Fisica "E.R.Caianiello", Universit\`a di Salerno, via Giovanni Paolo II 132, 84084 Fisciano (SA), Italy}
\address{$^2$ INFN, Gruppo collegato di Salerno, and CNISM, Unit\`a di Salerno, Universit\`a di Salerno, via Giovanni Paolo II 132, 84084 Fisciano (SA), Italy}

\ead{marryo36@hotmail.it}

\begin{abstract}
We study the consensus probability in Voter Model and Invasion Process starting from a seed initial configuration. In the case where the opinions have the same strength or slightly different (weak bias) this function was computed analytically by Sood, Antal and Redner and depends only on the degree of the promoter individual. We check numerically through large scale simulations the above mentioned theory and we find that in the case of strong bias a correlation between the consensus probability and other centrality measures emerge and Sood et al's theory is broken. 
\end{abstract}

\section{Introduction}

The manner in which opinions and ideas spread in a community is a topic of interest for the whole scientific community. 
From the beginning of the millennium scientists found that many real-world systems as the Worldwide Web or the Twitter Community can be well approximated by theoretical networks  such as the Scale-Free network~\cite{Bara2001, Alb, Aparicio}.
The latter one is a very heterogeneous network that well describes social media systems in which very few actors are very popular.
In order to understand opinion spreading from a theoretical point of view, consensus models on such graphs can be considered.
We will study toy models such as the Voter Model (VM) and the Invasion Process (IP)(see~\cite{CastSoc} for a review).
In these models a binary variable (0 or 1) is assigned to each node representing its opinion state.
The two models differ by their stochastic interaction rule as follows:
\begin{itemize}
\item in the VM a node $i$ is chosen randomly with probability $1/N$ where $N$ is the total number of nodes,
then a second node $j$ is picked among its neighbors, if $j$ has a different opinion then $i$ changes state.
\item in the IP $i$ is picked randomly in the network and then exports its state to one randomly chosen neighbor.
\end{itemize}
The dynamic process reaches an end when all nodes have the same opinion (\textit{consensus state}).
In regular graphs, where all nodes have the same number of connections $k$ (also called degree), these two models are equivalent.
In networks with an heterogeneous degree distribution this is no longer true and the two models have to be studied separately.
If an opinion state is favoured with respect to the other a \textit{bias parameter} $s$ is introduced (for further details see~\cite{Sood}).

\section{Sood et al's theory}
In the following we will consider an uncorrelated Scale-Free Network built using the \textit{uncorrelated configuration model} introduced by Catanzaro et al.~\cite{Catanzaro}. We will use networks with size $N=100$ and $N=200$.
Since we have checked that changing $N$ gives similar results, we 
used $N=100$ in cases when a larger statistical average was needed, and $N=200$
to consider a larger range of degrees of the nodes.
We will discuss only the most important results and leave the theorical details behind to the interested reader~\cite{Sood, CastSoc}.

\subsection{Unbiased Case}
Let us introduce the opinion state of node $i$, namely $\eta(i)$,
and the degree-weighted moments $\omega_m$
\begin{equation}
\omega_m = \frac{1}{N \langle k^m \rangle} \sum_i k_i ^m \eta(i).
\end{equation}
It can be shown~\cite{Sood, Suchecki} that in heterogeneous networks $\omega_1$ or $\omega_{-1}$ is conserved during the Voter Model process or the Invasion Process, respectively.
From these conserved quantities it is easily recognized that the consensus 1 probability 
$\varepsilon_1$ for the two unbiased processes is 
\begin{equation}
\varepsilon_1 ( \omega_1 ) = \omega_1
\label{vmueq}
\end{equation} 
for the VM, and
\begin{equation}
\varepsilon_1 ( \omega_{-1} ) = \omega_{-1}
\label{ipueq}
\end{equation}
for the IP.
Eqs.~(\ref{vmueq}) and~(\ref{ipueq}) are the main outcomes of the Sood et al's theory~\cite{Sood} for the unbiased case (i.e. $s=0$).
We will check numerically these results starting with a Seed Initial Configuration (SIC).
In SIC there is only 1 node with opinion 1 at the start (seed or promoting individual) of the process, then
\begin{equation}
\varepsilon_1 (k) = \frac{1}{N \langle k \rangle} k
\label{vmueq1}
\end{equation}
for the VM, and
\begin{equation}
\varepsilon_1 (k) = \frac{1}{N \langle k^{-1} \rangle} k^{-1}
\label{ipueq1}
\end{equation}
for the IP.

\begin{figure}[H]%
    \centering
    \subfloat[VM]{{\includegraphics[width=2.5in]{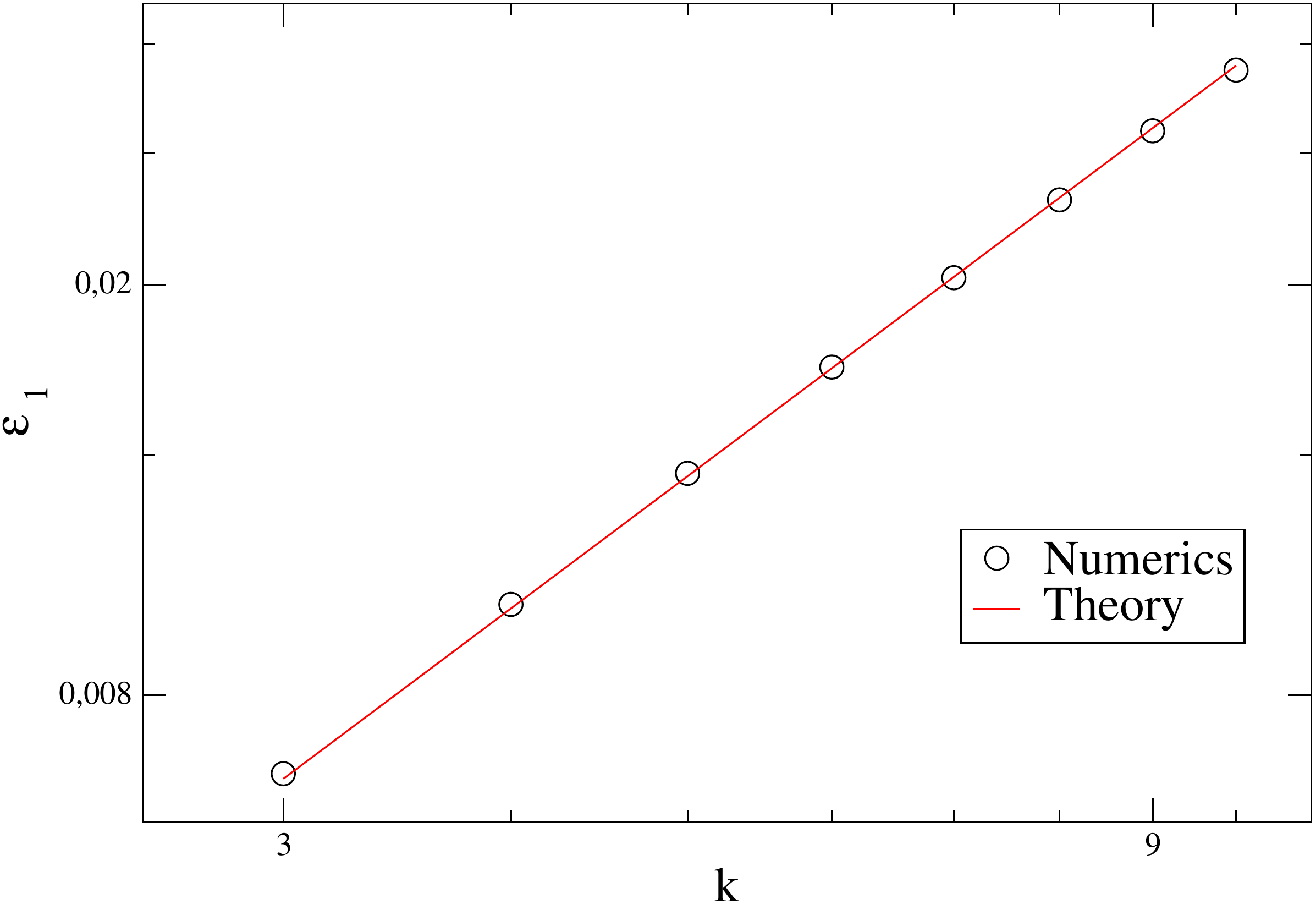} }}%
    \qquad
    \subfloat[IP]{{\includegraphics[width=2.5in]{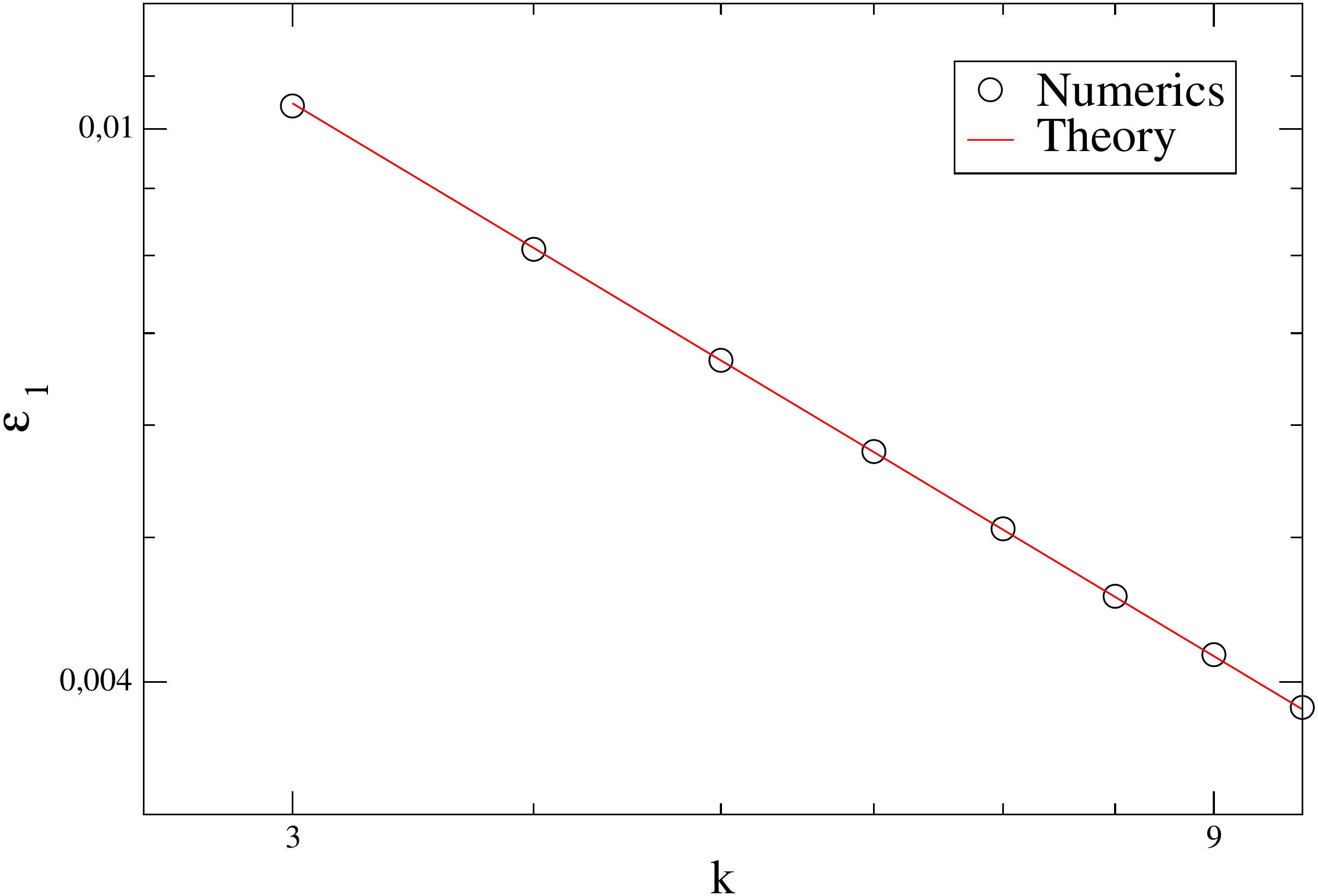} }}%
    \caption{$\varepsilon_1$ as a function of $k$ for unbiased VM (left panel) and IP (right panel) on a scale-free network with $N=100$. The numerical outcome is plotted with black circles and the theoretical curves of Eqs.~(\ref{vmueq1}) and~(\ref{ipueq1}) are plotted in red.}%
    \label{vmu100}%
\end{figure}

Let us now discuss our numerical results.
In Fig.~\ref{vmu100} we plot $\varepsilon_1 (k)$ for a scale-free network with $N=100$. The relative error between theory and numerical data is $0.5\%$  
for the VM and $0.3 \%$ for the IP.

\subsection{Biased Case}
In the biased case (i.e. for $s>0$) Sood et al. find the analytical form of $\varepsilon_1$ using the annealed approximation~\cite{Sood, Satorras} in the weak selection limit $s \ll 1$. They find a backward Kolmogorov equation for $\varepsilon_1$ of the form
\begin{equation}
\omega_1 ( 1 - \omega_1 ) \left[ s \frac{\partial}{\partial \omega_1} + \frac{\langle k^2 \rangle}{\langle k \rangle ^2 N} \frac{\partial^2}{\partial \omega_1 ^2} \right] \varepsilon_1 (\omega_1) = 0
\end{equation}
for the biased VM, and
\begin{equation}
\frac{\omega_{-1} ( 1 - \omega_{-1} )}{\langle k \rangle \langle k^{-1} \rangle} \left[ s \frac{\partial}{\partial \omega_{-1}} + \frac{1}{N} \frac{\partial^2}{\partial \omega_{-1} ^2} \right] \varepsilon_1 (\omega_{-1}) = 0
\end{equation}
for the biased IP.

\begin{figure}[H]%
    \centering
    \subfloat[VM]{{\includegraphics[width=2.5in]{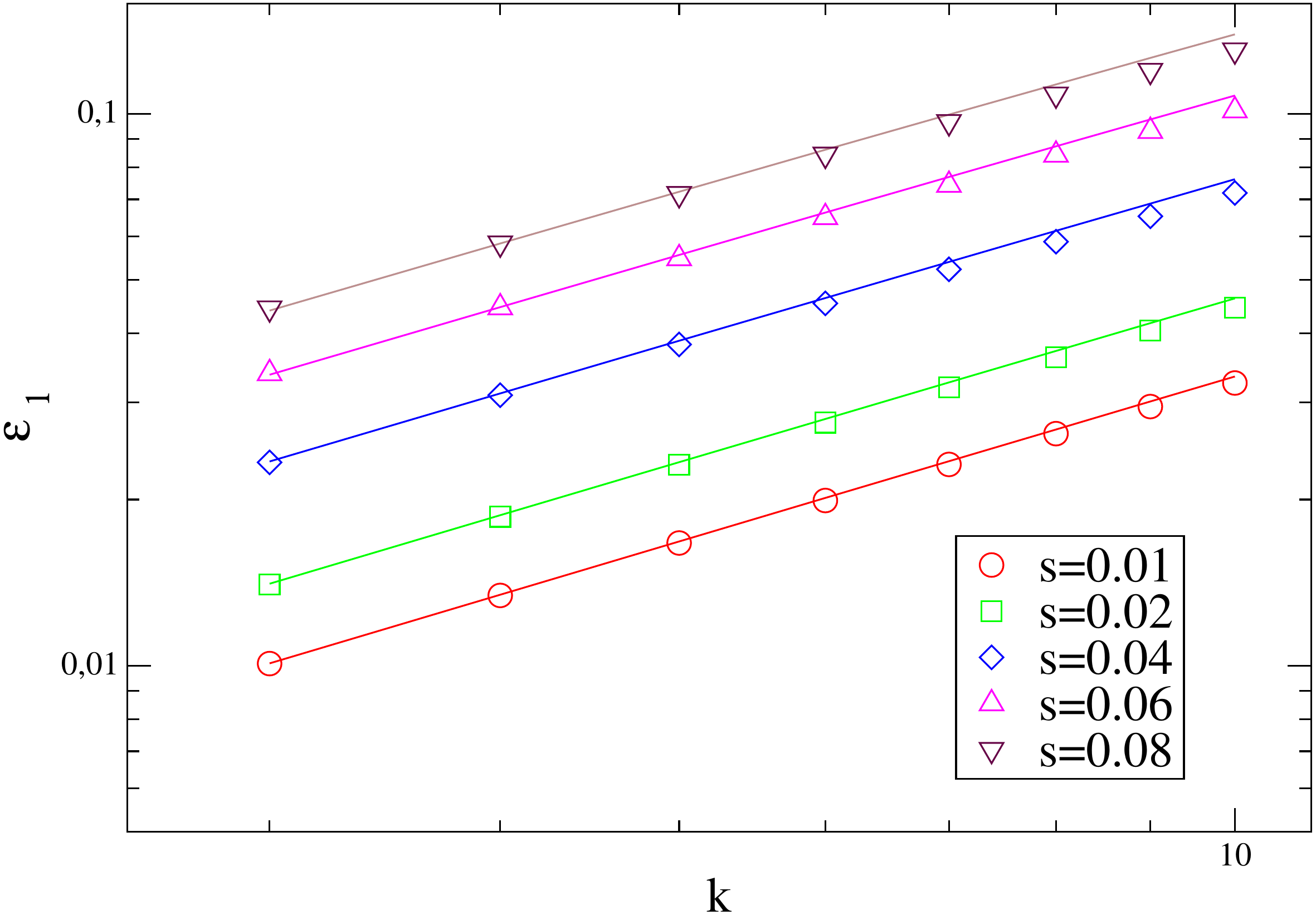} }}%
    \qquad
    \subfloat[IP]{{\includegraphics[width=2.5in]{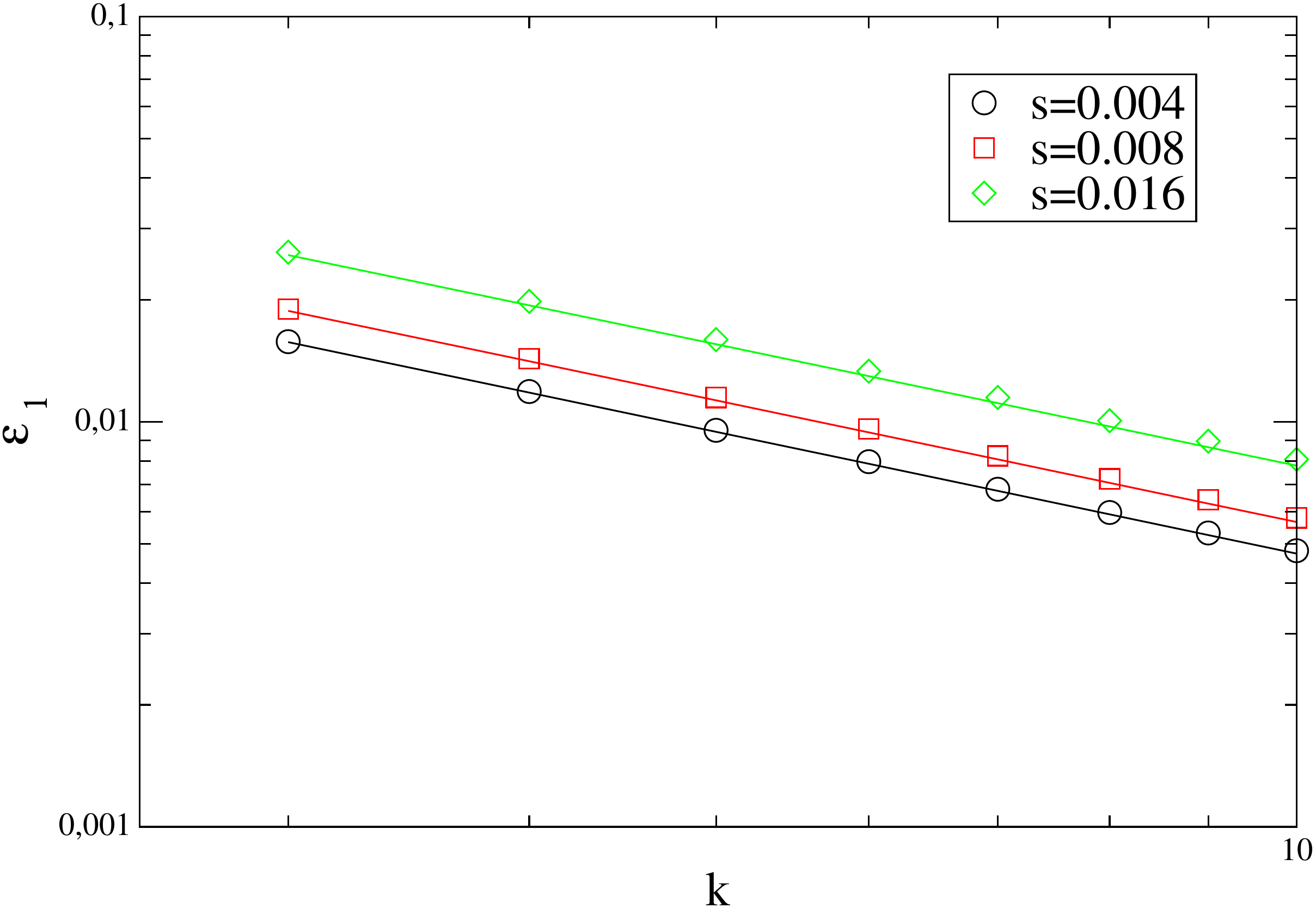} }}%
    \caption{$\varepsilon_1$ as a function of $k$ in the biased VM (left panel) and IP (right panel) for different values of the bias $s$ with $N_d= 10^6$ stochastic realizations.
Different colors correspond to different values of $s$. The process is simulated on a single realization of a scale-free with $N=100$. The theoretical curves of Eqs.~(\ref{vmsred}) and~(\ref{ipsred}) are plotted with lines, the numerical outcomes are plotted with symbols.}%
    \label{vms100}%
\end{figure}

These equations are very similar to those of a Random Walker with absorbing boundaries whose solution for the VM is
\begin{equation}
\varepsilon_1 ( \omega_1 ) = \frac{1-e^{-sN_{eff} \omega_1}}{1-e^{-sN_{eff}}} 
\label{vmsred}
\end{equation}
where $N_{eff} = N \frac{\langle k \rangle ^2}{\langle k^2 \rangle}$, while for the IP it is
\begin{equation}
\varepsilon_1 ( \omega_{-1}) = \frac{1-e^{-sN_{eff} \omega_{-1}}}{1-e^{-sN_{eff}}}
\label{ipsred}
\end{equation}
with $N_{eff} = N \langle k \rangle \langle k^{-1}\rangle$.
In Fig.~\ref{vms100} we compare the theoretical functional form of $\varepsilon_1$ given by Eqs.~(\ref{vmsred}) and~(\ref{ipsred}) to the results of our simulations. The relative error between theory and numerical data is around $1-3\%$ both for the VM and for the IP, increasing with increasing $s$ and $k$.
We checked the nature of the discrepancies in increasing $k$ varying the size of the network for fixed $s$ and we found that for fixed $k$ the error decreases increasing $N$, thus they are due to finite size effects.
Let us note that in Fig.~\ref{vms100} we used different values of $s$ for the two models. Sood et al. inferred that the weakly bias regime is given by $1/N_{eff} \ll s \ll 1$. For the analysed cases $(N_{eff})_{VM} < (N_{eff})_{IP}$ and so it is reasonable to use greater values of $s$ for the VM. 

\subsection{Strong bias}
\begin{figure}[H]%
    \centering
    \subfloat[VM]{{\includegraphics[width=2.5in]{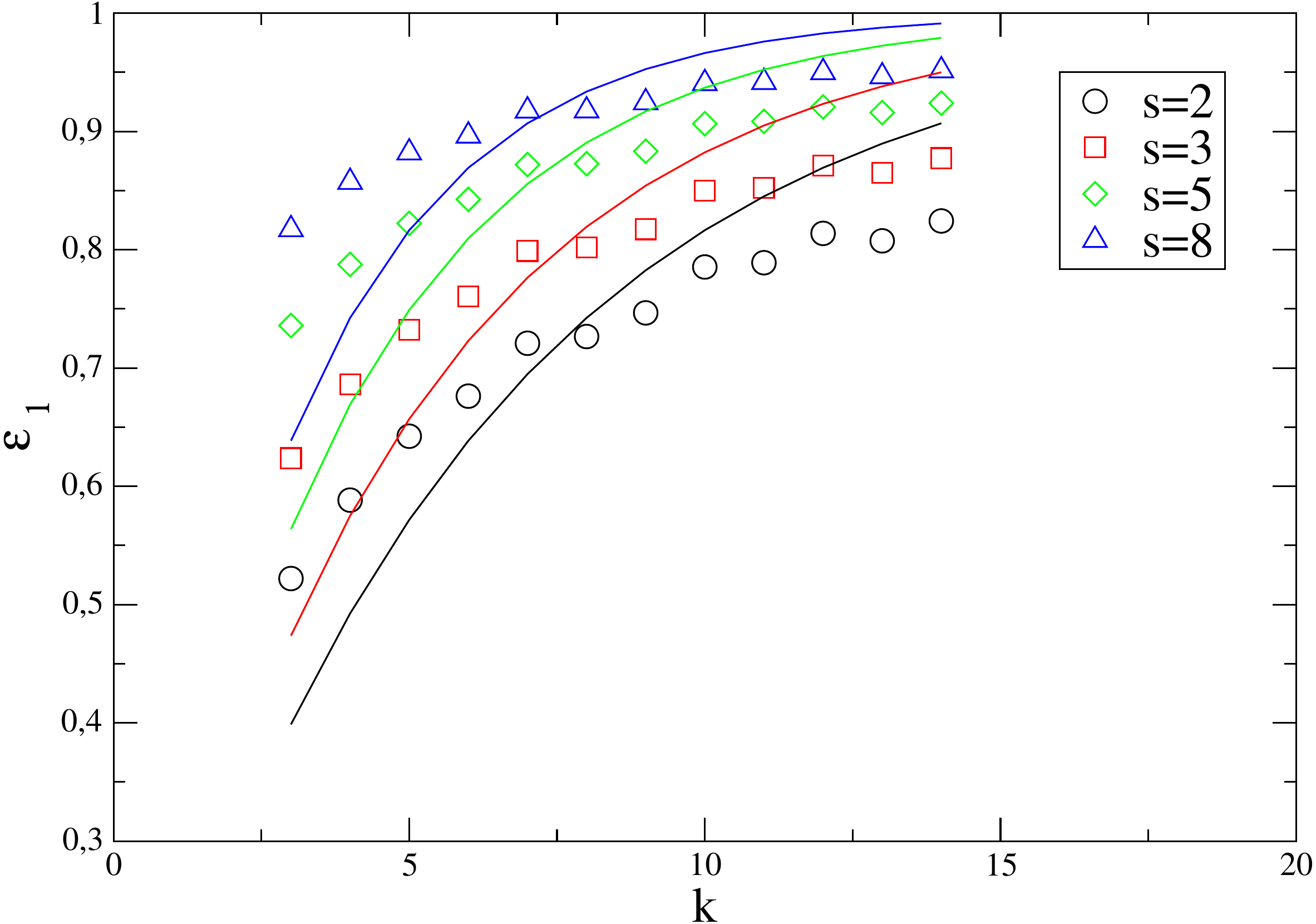} }}%
    \qquad
    \subfloat[IP]{{\includegraphics[width=2.5in]{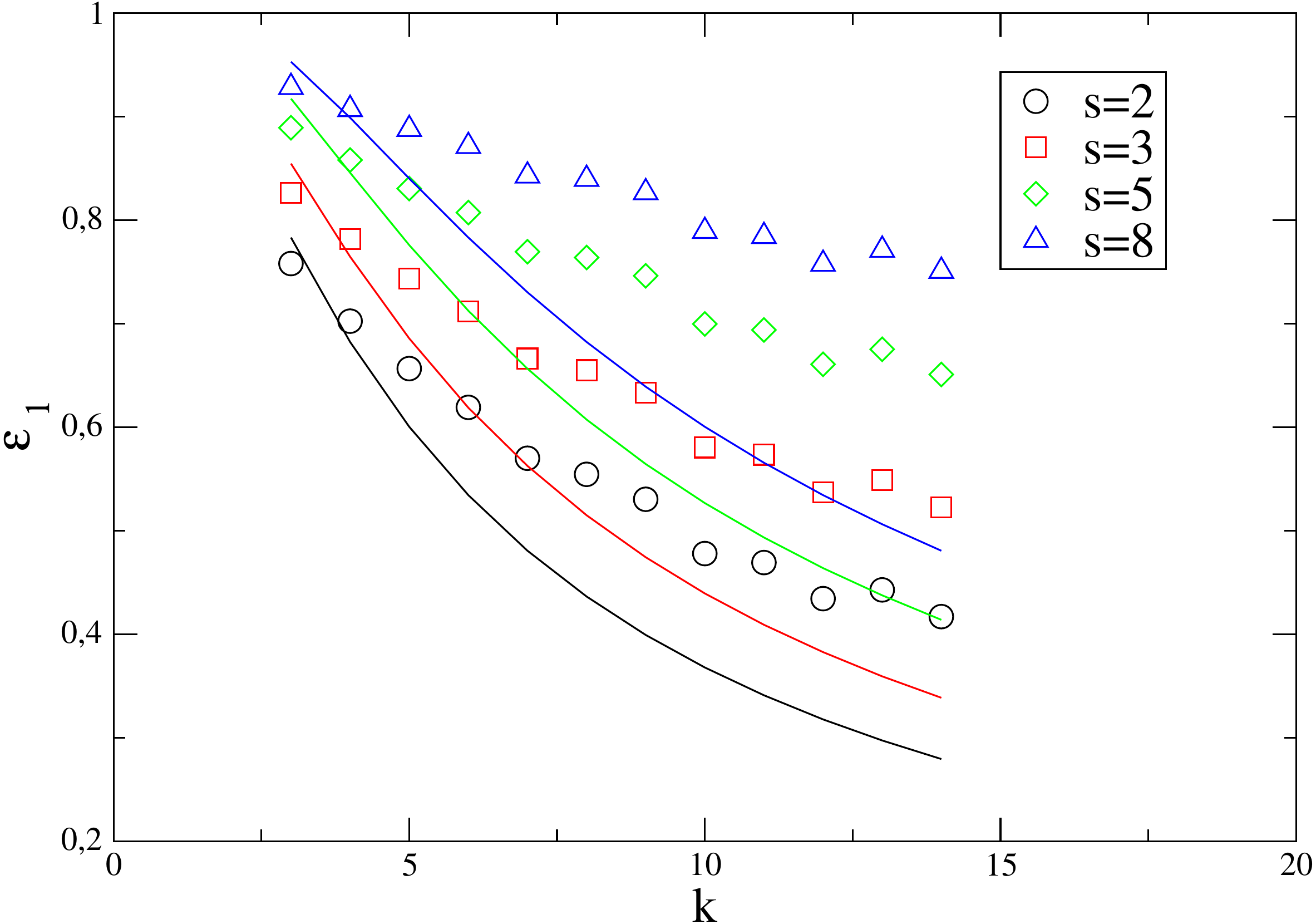} }}%
    \caption{$\varepsilon_1$ as a function of $k$  for the biased VM (left panel) and IP (right panel) for different high values of the bias $s$, for $N_d=10^6$ stochastic realizations. Different symbols correspond to different values of $s$. The process is simulated on a single realization of a scale-free network with $N=200$. The numerical outcomes are plotted with circles and the Sood's curves of Eqs.~(\ref{vmsred}) and~(\ref{ipsred}) are plotted with lines.}%
    \label{Hvms}%
\end{figure}

We denote as \textit{strong bias} the regime in which $s \gtrsim 1 $. In this case we can see from Fig.~\ref{Hvms} that Sood et al's theory (plotted with lines) is not in agreement with simulation data (symbols).

\begin{figure}[H]%
    \centering
    \subfloat[VM weak bias]{{\includegraphics[width=2.5in]{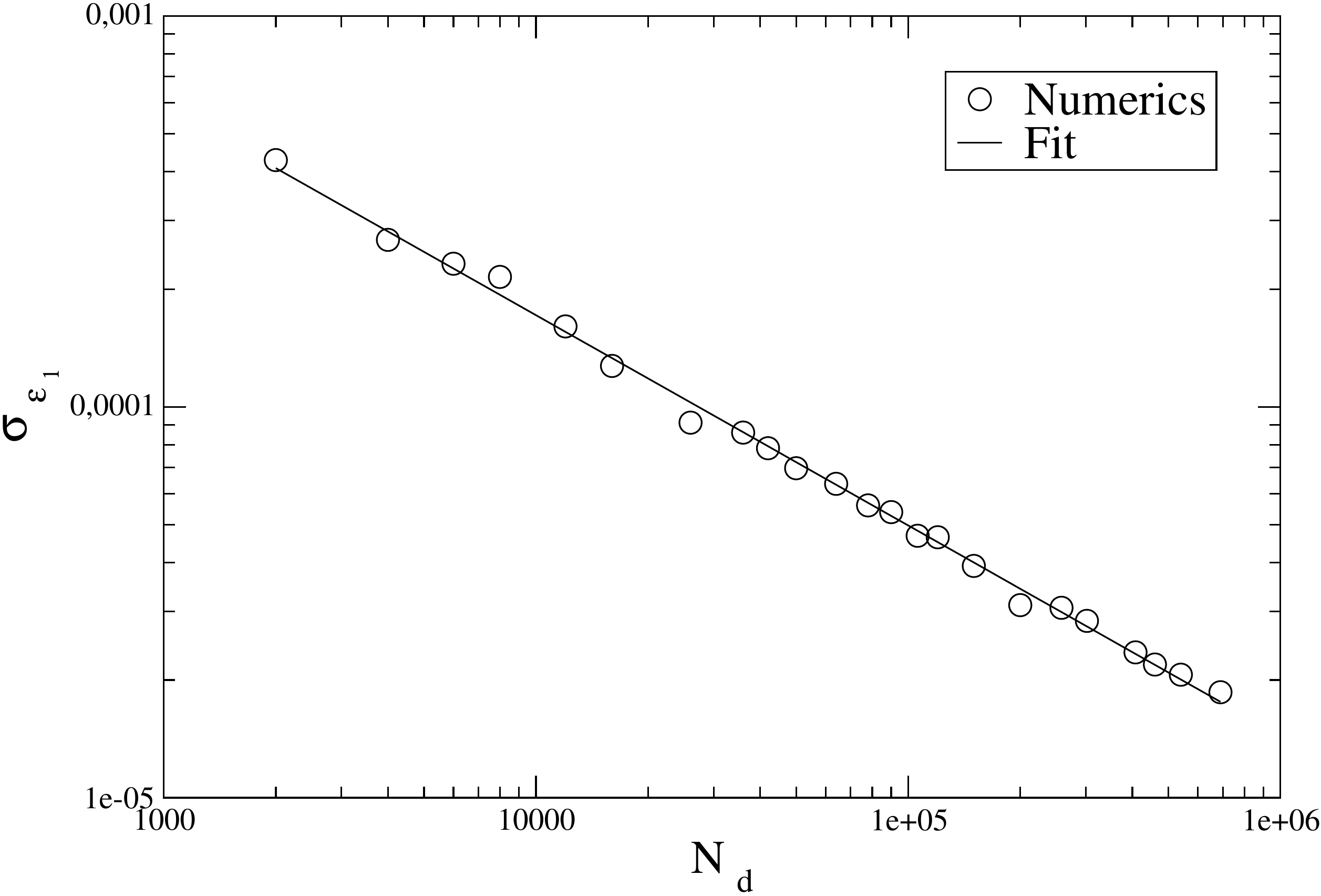} }}%
    \qquad
    \subfloat[VM strong bias]{{\includegraphics[width=2.5in]{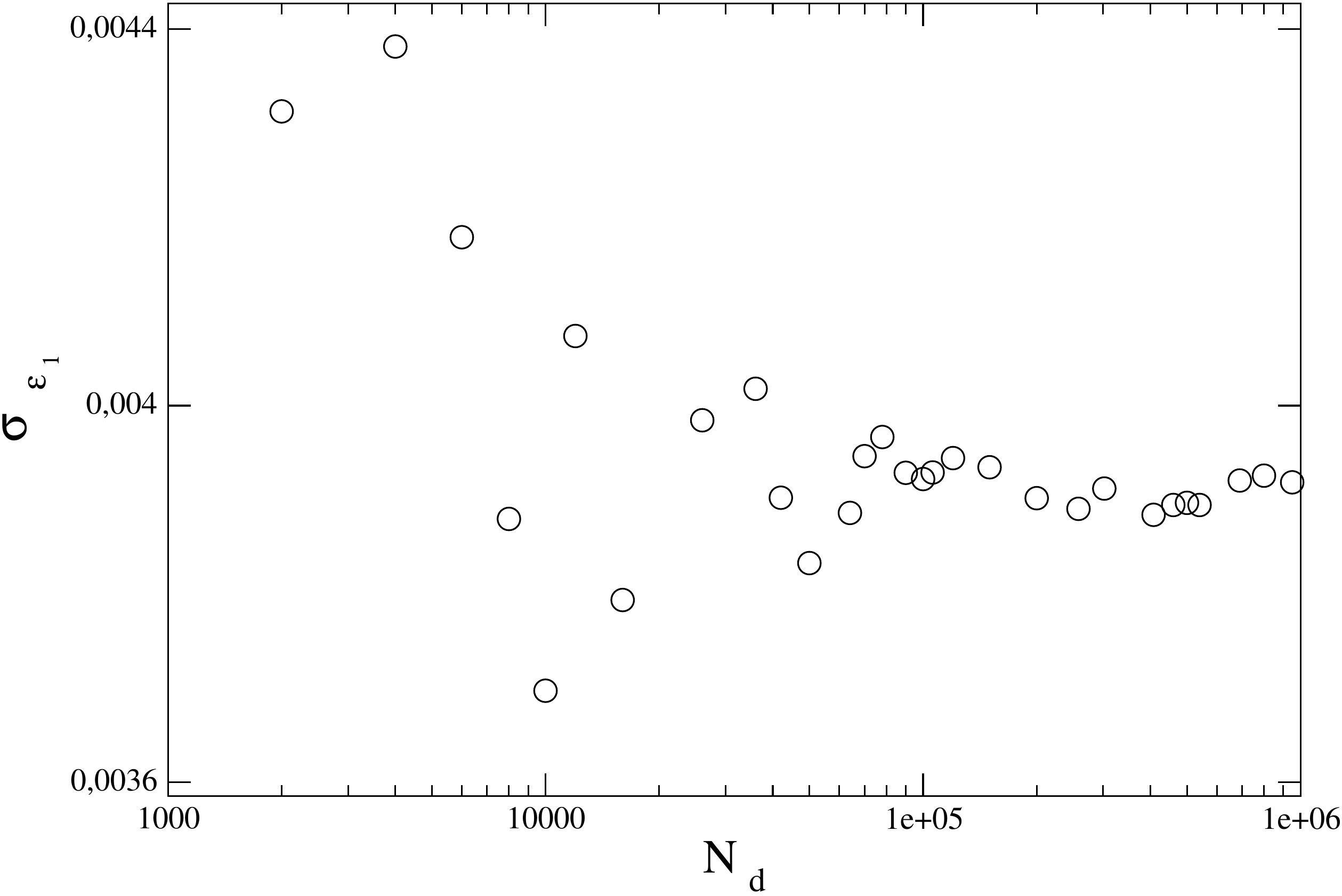} }}%
    \caption{$\sigma_{\varepsilon_1}$ as a function of $N_d$ for the nodes with $k=3$ of the biased VM with $s=0.01$ (left panel) and with $s=3$ (right panel). The process is simulated on a single realization of a scale-free network with $N=100$. The numerical outcome is plotted with black circles, the best fit of Eq.~(\ref{clth}) is plotted with a black line.}%
    \label{devms1}%
\end{figure}

Let us study the standard deviation $\sigma_{\varepsilon_1}$ of $\varepsilon_1$. If $\varepsilon_1$ depends only on $k$ then $\sigma_{\varepsilon_1}$ should scale as $\frac{1}{\sqrt{N_d}}$, where $N_d$ is the number of stochastic realizations used in the Montecarlo simulation, according to the central limit theorem.
This is valid for $s \ll 1$ as we can see in Fig.~\ref{devms1}a where $\sigma_{\varepsilon_1}$ is well fitted by
\begin{equation}
\sigma_{\varepsilon_1} \simeq N_d ^{-\beta}
\label{clth}
\end{equation}
with $\beta \approx 0.5$. However the central limit theorem is not obeyed for $s\gtrsim 1$ as we can see in Fig.\ref{devms1}b.
For IP we find similar results.

\section{Emergence of correlations between $\varepsilon_1$ and other topological measures}
The fact that, for large $s$, $\varepsilon_1$ does not obey the central limit theorem suggests that it may depend on other quantities beside $k$.
For this reason we study the possible correlations with other topological measures as the average neighbor's degree $k_{nn}$ and the eigenvector centrality $x$ (for further details on centrality measures see~\cite{Barab}).
The eigenvector centrality has been used largely in the literature for quantifying the dynamic influence of a node~\cite{klemm}.

\begin{figure}[H]%
    \centering
    \subfloat[VM weak $s$]{{\includegraphics[width=2.5in]{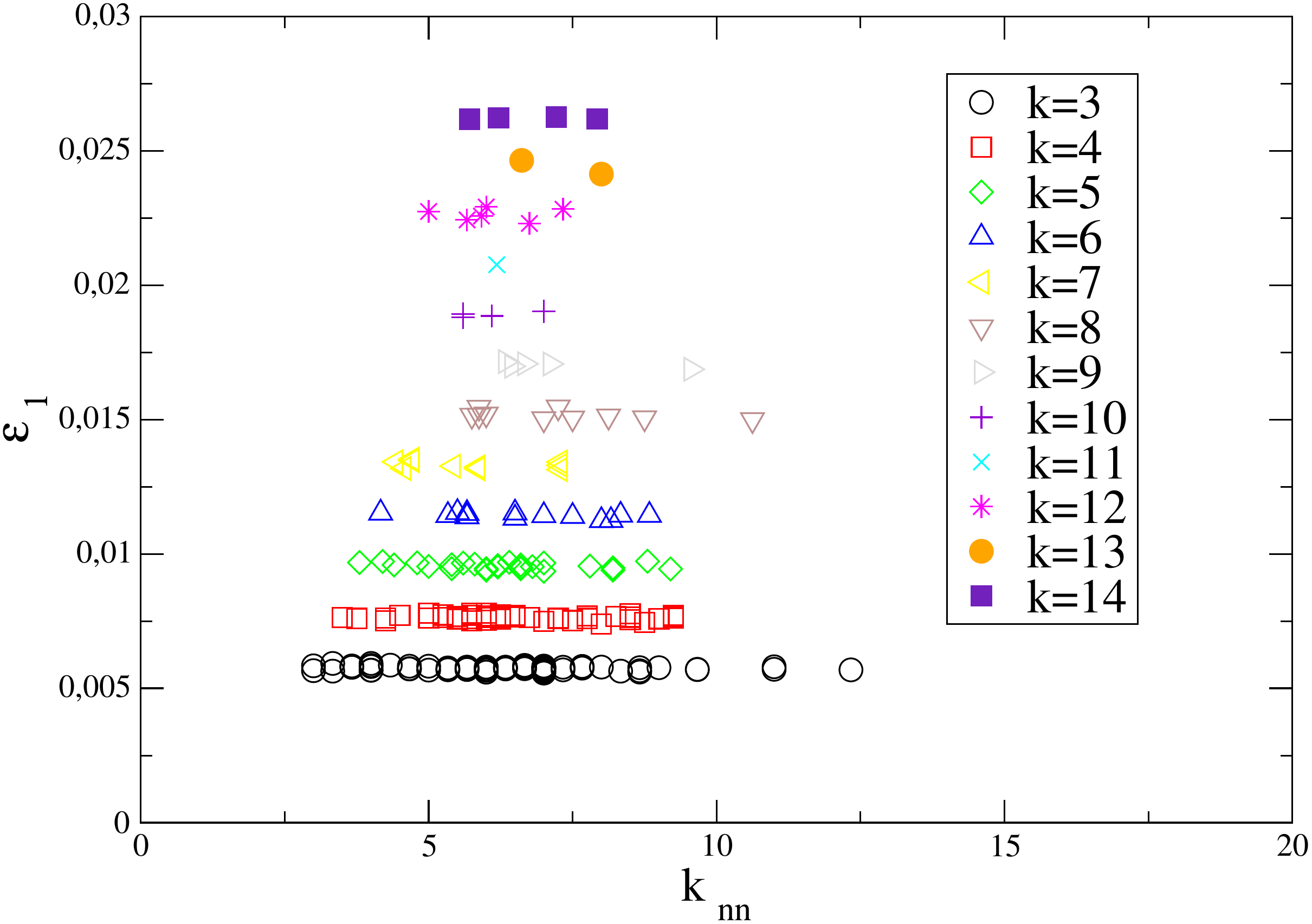} }}%
    \qquad
    \subfloat[VM strong $s$]{{\includegraphics[width=2.5in]{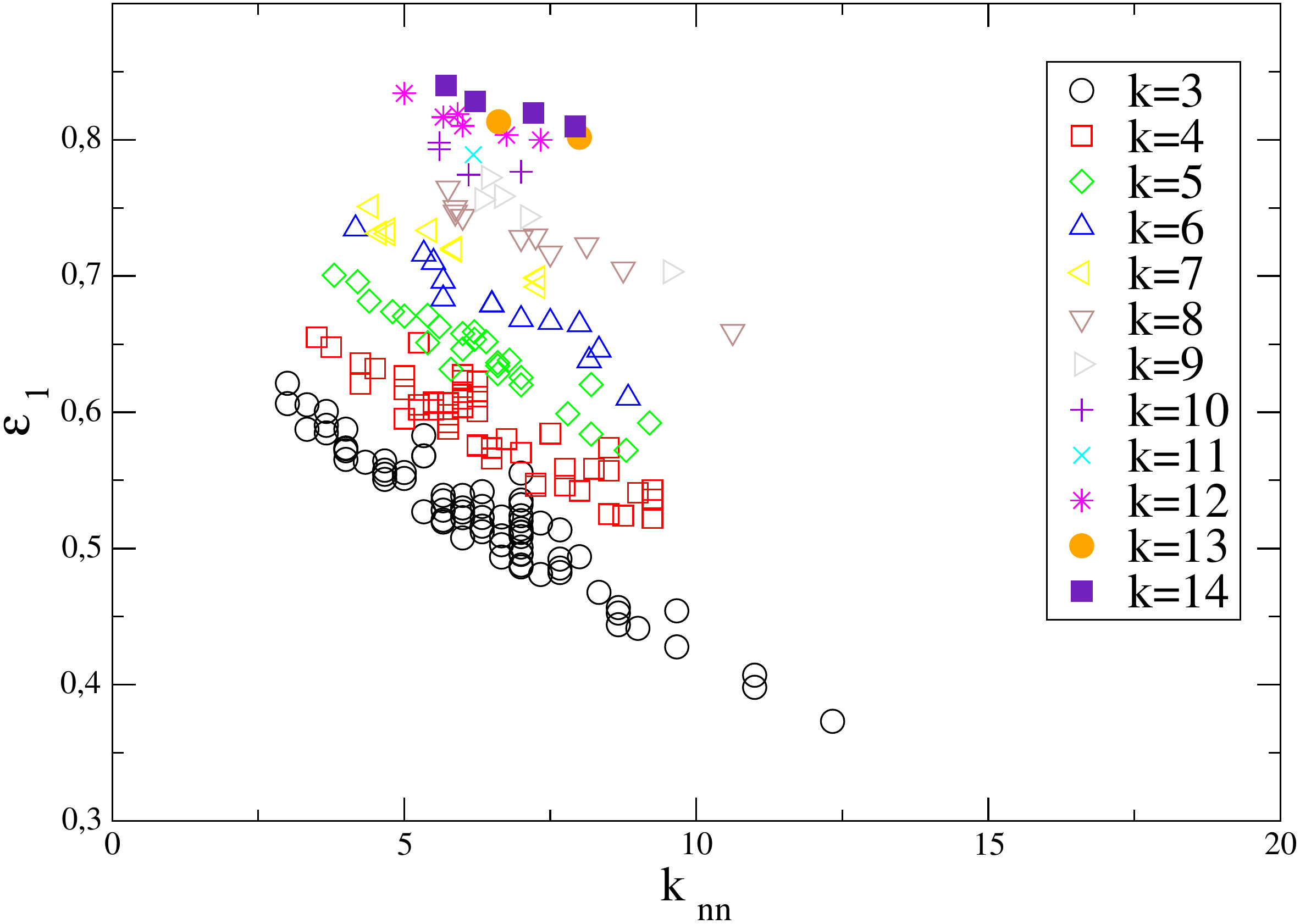} }}%
    \caption{$\varepsilon_1$ against $k_{nn}$ for a biased VM with $s=0.01$ (left panel) and $s=3$ (right panel). Different symbols correspond to different values of $k$ (see key). The process is simulated for $N_d=10^6$ stochastic realizations of the dynamics on a single realization of a scale-free network with $N=200$.}%
\label{knnvms}%
\end{figure}

Regarding $k_{nn}$, for the biased VM, when we increase $s$ from the weak to the strong regime, we go from an uncorrelated behavior to an anticorrelated one for fixed $k$, as it can be seen in Fig.~\ref{knnvms}.

\begin{figure}[H]%
    \centering
    \subfloat[IP weak $s$]{{\includegraphics[width=2.5in]{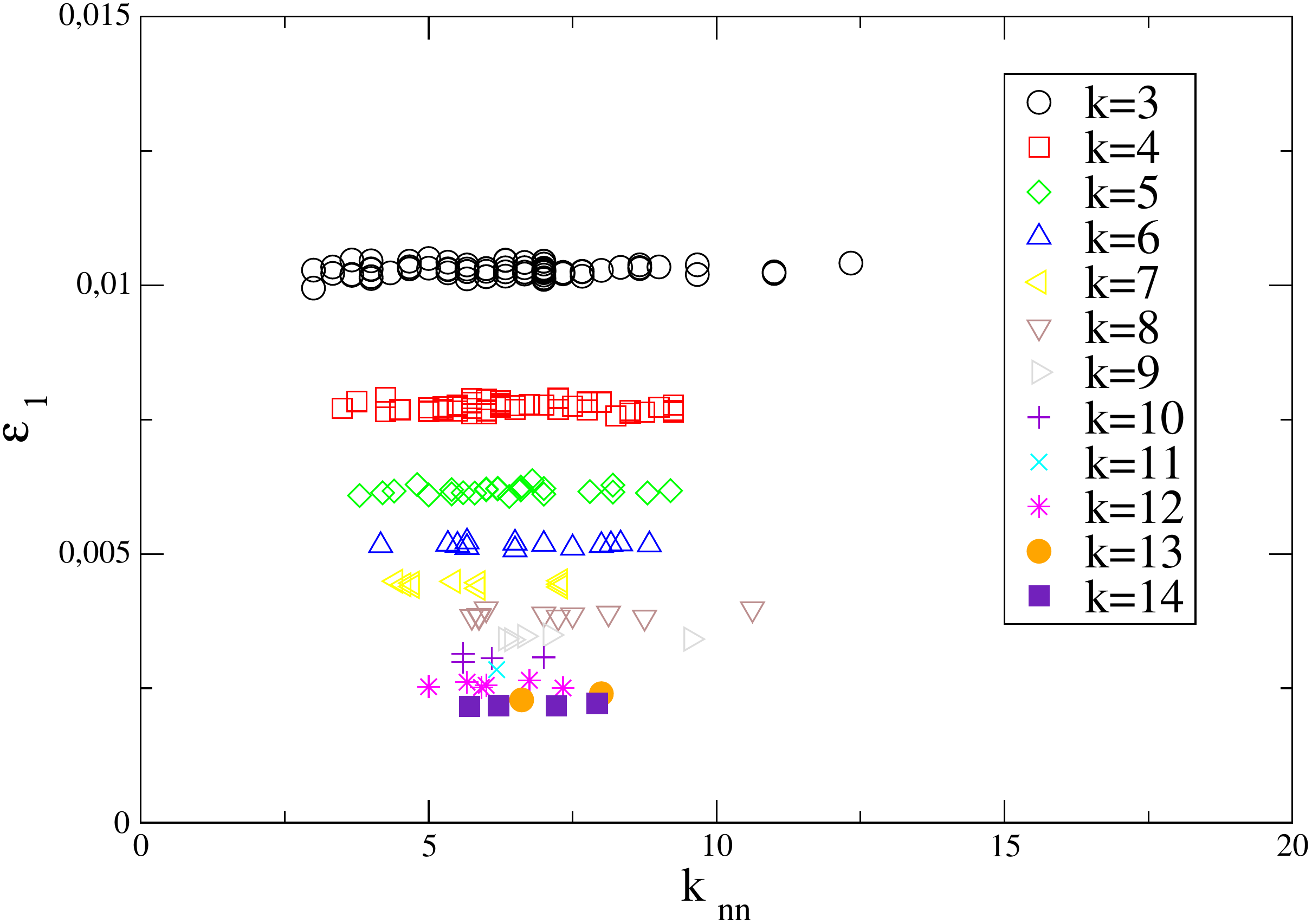} }}%
    \qquad
    \subfloat[IP strong $s$]{{\includegraphics[width=2.5in]{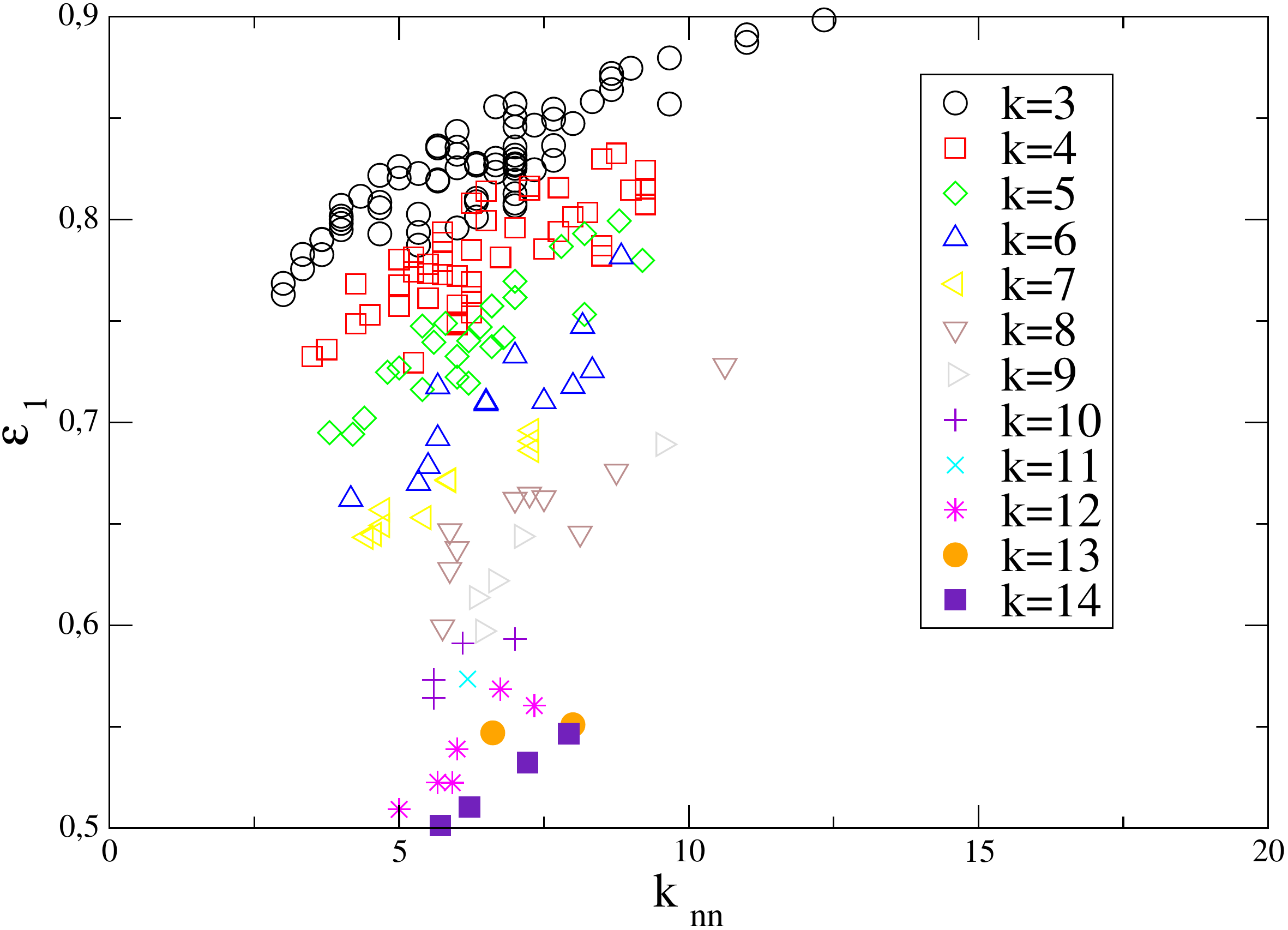} }}%
    \caption{$\varepsilon_1$ against $k_{nn}$ for a biased IP with $s=0.004$ (left panel) and $s=3$ (right panel). Different symbols correspond to different values of $k$ (see key). The process is simulated for $N_d=10^6$ stochastic realizations of the dynamics on a single realization of a scale-free network with $N=200$.}%
\label{knnips}%
\end{figure}

Instead, for the biased IP, we see in Fig.~\ref{knnips} that setting $s$ in the strong bias regime induces correlations effects between $\varepsilon_1$ and $k_{nn}$.

The correlations between $\varepsilon_1$ and the eigenvector centrality $x$ are qualitatively similar to those with $k_{nn}$.

\section{Conclusions}
We studied the behavior of VM and IP with a null, weak and strong bias in seed initial configuration.
While for null and weak bias the solutions given by Sood et al's theory are accurate, for strong bias they fail to predict the form of $\varepsilon_1$.
We prove that this is due to the emergence of correlations between $\varepsilon_1$ and other topological measures such as $k_{nn}$ and $x$ in the biased VM or biased IP. Then we conclude that $\varepsilon_1$ does not depend only on $k$ and a mathematical formulation that takes into account the presence of other correlations is needed.

\ack
We warmly thank Dr. C. Castellano for discussions.


\section*{References}
\begin{thebibliography}{9}
\bibitem{Bara2001}
\hspace{5mm}Barab\`asi A L and Albert R 2002 \textit{Rev. Mod. Phys.} \textbf{74} 47
\bibitem{Alb}
\hspace{5mm}Albert R, Jeong H and Barab\`asi A L 1999 \textit{Nature} \textbf{401} 130-131
\bibitem{Aparicio}
\hspace{5mm}Aparicio S, Villaz\'on-Terrazas J and \'Alvarez G 2015 \textit{Entropy} \textbf{17} 5848-5867 
\bibitem{CastSoc}
\hspace{5mm}Castellano C, Fortunato S and Loreto V 2009 \textit{Rev. Mod. Phys.} \textbf{81} 591
\bibitem{Sood}
\hspace{5mm}Sood V, Andal T and Redner S 2008 \textit{Phys. Rev. E.} \textbf{77} 041121
\bibitem{Catanzaro}
\hspace{5mm}Catanzaro M, Bogu\~n\'a M and Pastor-Satorras R 2004 \textit{Phys. Rev. E.}\textbf{71} 027103
\bibitem{Suchecki}
\hspace{5mm}Suchecki K, Eguiluz V M  and San Miguel M 2005 \textit{Europhys. Lett.} \textbf{69} (2) 228-234 
\bibitem{Satorras}
\hspace{5mm}Pastor-Satorras R, Castellano C, Mieghem P V, Vespignani A 2006  \textit{Physica A} \textbf{363} 1
\bibitem{Barab}
\hspace{5mm}Barab\`asi A L 2016 \textit{Network Science}, Cambridge University Press
\bibitem{klemm}
\hspace{5mm}Klemm K, Serrano  M A, Eguiluz V M and San Miguel M 2012  \textit{Sci.Rep.} \textbf{2}: 292

\end{thebibliography}
\end{document}